\documentclass[sn-nature,pdflatex]{sn-jnl}% Style for submissions to Nature Portfolio journals

\usepackage{lmodern}
\usepackage{anyfontsize}
\usepackage{fix-cm}
\usepackage{multirow}%
\usepackage{amsmath,amssymb,amsfonts}%
\usepackage{amsthm}%
\usepackage{mathrsfs}%
\usepackage[title]{appendix}%
\usepackage{xcolor}%
\usepackage{textcomp}%
\usepackage{manyfoot}%
\usepackage{booktabs}%
\usepackage{algorithm}%
\usepackage{algorithmicx}%
\usepackage{algpseudocode}%
\usepackage{listings}%
\usepackage{hyperref}
\usepackage{adjustbox}
\usepackage{float}
\usepackage{lineno}

\usepackage{graphicx}
\usepackage[english]{babel}

\definecolor{codegreen}{rgb}{0,0.6,0}
\definecolor{codegray}{rgb}{0.5,0.5,0.5}
\definecolor{codepurple}{rgb}{0.58,0,0.82}
\definecolor{backcolour}{rgb}{0.95,0.95,0.92}

\renewcommand{\cite}[1]{\textsuperscript{\citenum{#1}}}
\lstdefinelanguage{markdown}{
    morekeywords={\#, \#\#, \#\#\#, \#\#\#\#, -, *, >, `, ```}, 
    sensitive=true, 
    morecomment=[l]{//}, 
    morecomment=[s]{/*}{*/}, 
    morestring=[b]", 
    morestring=[b]', 
}
\lstdefinestyle{markdownstyle}{
    language=markdown,
    basicstyle=\ttfamily\footnotesize,
    keywordstyle=\color{blue}\bfseries,
    commentstyle=\color{gray}\ttfamily,
    stringstyle=\color{red}\ttfamily,
    showstringspaces=false,
    breaklines=true,
    frame=single,
    backgroundcolor=\color{gray!10}
    }

\lstdefinestyle{mystyle}{
    backgroundcolor=\color{backcolour},   
    commentstyle=\color{codegreen},
    keywordstyle=\color{black},
    numberstyle=\tiny\color{black},
    stringstyle=\color{codegreen},
    basicstyle=\ttfamily\footnotesize,
    breakatwhitespace=false,         
    breaklines=true,                 
    captionpos=b,                    
    keepspaces=true,                                   
    numbersep=5pt,                  
    showspaces=false,                
    showstringspaces=false,
    showtabs=false,                  
    tabsize=2,
    language=Python
}

\definecolor{darkgreen}{rgb}{0,0.5,0}

\newcommand{\revise}[1]{\textcolor{darkgreen}{[#1]}}

\title{StarWhisper Telescope: An AI framework for automating end-to-end astronomical observations}

\author{%
  \textbf{Cunshi Wang}\affilone\affiltwo\footnote{Email:\url{wangcunshi@nao.cas.cn}},
  \textbf{Yu Zhang}\affiltwo,
  \textbf{Yuyang Li}\affilone\affiltwo\footnote{Email:\url{liyuyang22@mails.ucas.ac.cn}},
  \textbf{Xinjie Hu}\affiltwo\affilthree,
  \textbf{Yiming Mao}\affilone\affiltwo\affilfour,
  \textbf{Xunhao Chen}\affilone\affiltwo,
  \textbf{Pengliang Du}\affiltwo,
  \textbf{Rui Wang}\affiltwo,
  \textbf{Ying Wu}\affiltwo\affilfive,
  \textbf{Hang Yang}\affilone\affiltwo,
  \textbf{Yansong Li}\affilsix,
  \textbf{Beichuan Wang}\affilone\affiltwo,
  \textbf{Haiyang Mu}\affiltwo,
  \textbf{Xiaohan Chen}\affilone\affiltwo,
  \textbf{Shunxuan He}\affilone\affiltwo,
  \textbf{Hao Mo}\affilone\affiltwo,
  \textbf{Liyue Zhang}\affilone\affiltwo,
  \textbf{Lin Du}\affilone\affiltwo,
  \textbf{Yunning Zhao}\affilone\affiltwo,
  \textbf{Jianfeng Tian}\affiltwo,
  \textbf{Liang Ge}\affiltwo\affilfive,
  \textbf{Yongna Mao}\affilone,
  \textbf{Shengming Li}\affiltwo,
  \textbf{Zheng Wang}\affiltwo,
  \textbf{Xiaomeng Lu}\affiltwo,
  \textbf{Jinhang Zou}\affiltwo,
  \textbf{Yang Huang}\affilone\affiltwo,
  \textbf{Ningchen Sun}\affilone\affiltwo,
  \textbf{Jie Zheng}\affiltwo,
  \textbf{Min He}\affiltwo,
  \textbf{Yu Bai}\affiltwo\affilfive,
  \textbf{Junjie Jin}\affiltwo,
  \textbf{Hong Wu}\affilone\affiltwo,
  \textbf{Jifeng Liu}\affilone\affiltwo\affilfive \\
  \affilone University of Chinese Academy of Sciences, Beijing, China;\\
  \affiltwo Key Laboratory of Optical Astronomy,\\ National Astronomical Observatories, Chinese Academy of Sciences, Beijing, China; \\
  \affilthree School of Computing Science, Simon Fraser University, Burnaby, British Columbia V5A1S6, Canada; \\
  \affilfour College of Science, Tibet University, Tibet, China; \\
  \affilfive Institute for Frontiers in Astronomy and Astrophysics,\\ Beijing Normal University, Beijing, China; \\
  \affilsix Liii Network Co., Zhejiang, China;
}

\newcommand{\affilone}{\textsuperscript{1}}
\newcommand{\affiltwo}{\textsuperscript{2}}
\newcommand{\affilthree}{\textsuperscript{3}}
\newcommand{\affilfour}{\textsuperscript{4}}
\newcommand{\affilfive}{\textsuperscript{5}}
\newcommand{\affilsix}{\textsuperscript{6}}

\begin{document}
%\linenumbers
% \blfootnote{\corrauth Corresponding authors}
\maketitle

\begin{abstract}

    The exponential growth of large-scale telescope arrays has boosted time-domain astronomy development but introduced operational bottlenecks, including labor-intensive observation planning, data processing, and real-time decision-making. Here we present the StarWhisper Telescope system, an AI agent framework automating end-to-end astronomical observations for surveys like the Nearby Galaxy Supernovae Survey. By integrating large language models with specialized function calls and modular workflows, StarWhisper Telescope autonomously generates site-specific observation lists, executes real-time image analysis via pipelines, and dynamically triggers follow-up proposals upon transient detection. The system reduces human intervention through automated observation planning, telescope controlling and data processing, while enabling seamless collaboration between amateur and professional astronomers. Deployed across Nearby Galaxy Supernovae Survey’s network of 10 amateur telescopes, StarWhisper Telescope has detected transients with promising response times relative to existing surveys. Furthermore, StarWhisper Telescope’s scalable agent architecture provides a blueprint for future facilities like the Global Open Transient Telescope Array, where AI-driven autonomy will be critical for managing 60 telescopes.
    %By unifying automated instrumentation with scientific reasoning, this work establishes a transformative paradigm where embodied AI systems accelerate discovery, setting a new benchmark for efficiency and scalability in astrophysical research.  (e.g., SN2024xin, AT2025pk)

\end{abstract}

%%%

%    In the field of astronomical observation, \revise{the soon-to-be-built large telescope arrays will significantly increase astronomers' workloads. Studies have shown that Large Language Model (LLM)-based agents can promote unmanned observation to alleviate labor-intensive tasks and reduce the costs associated with training personnel.}
    %
%    In this paper, we introduce the \textbf{StarWhisper Telescope System}, developed for the Nearby Galaxy Supernovae Survey (NGSS) project. NGSS is a large-scale astronomical survey aimed at discovering transient, i.e., supernovae, in nearby galaxies. The observation process—including generating observation lists, conducting observations, analyzing data, and providing feedback—is fully automated by \revise{the designed AI agents system.}
    %
%    Observation lists are tailored for different sites and strategies to ensure comprehensive coverage of celestial objects. After manual verification, these lists are uploaded to the telescopes by system agents, initiating observations through natural language commands.
    %
 %   Captured images are analyzed in real-time, and detected transients are immediately reported to observers. The agent converts these reports into real-time follow-up observation proposals, which are sent to the Xinglong Observatory group chat and added to the next-day observation lists.
    %
%    Moreover, the integration of AI agents allows online system access, saving astronomers valuable time and encouraging greater participation from amateur astronomers in the NGSS project.

\section{Introduction}\label{sec:intro}

As time domain astronomy develops, large scientific facilities are in the planning. The Global Open Transient Telescope Array (GOTTA~\cite{2021Sitian, huang2025minisitianarraypathfindersitian}, which also referred to the SiTian project.), aims to construct 60 1-meter telescopes in its first stage to build an array to monitor the sky. The labor cost for the observation of the GOTTA is huge, with an estimated observation personnel sequence exceeding 200 people. Most of the telescopes will be deployed at the Lenghu Observatory, with a latitude of about 4,200 meters. Due to the large number of telescopes, automated observations are 
necessary for future time-domain surveys. With the help of a Large Language Model (LLM) based agent, new automated observations will aid astronomers by taking the observations and processing the data, realizing the concept of an "AI Astronomer".

Astronomical observation can be broadly divided into three phases: (1) observation planning, (2) observation execution, and (3) data processing. These phases are executed sequentially, starting with planning, followed by execution, and concluding with data processing. Notably, the results of data processing feed back into the planning phase, influencing future observation strategies and helping to optimize the overall observational process.

The observation planning phase aims to create an observation list for each telescope participating in the night's observations. This list includes the observation time (in UTC time), the astronomical objects to be observed, their coordinates (typically in Right Ascension, R.A., and Declination, Dec.), and other relevant information.
Observation control is typically managed by the Observation Control System (OCS, \cite{Wang2024mSTOCS}), which operates based on the provided observation list and additional parameters, such as the exposure time, number of exposures, required filters,
telescope position and focus control. The images observed by the telescope are subsequently transferred to the data pipeline~\cite{Li2024mSTpipeline} for processing, extracting the photometric information required by astronomers, and identifying transients as high-value objects.

The observation, planning, and data processing phases are labor-intensive and require considerable effort.
Generating the observation list involves selecting the most suitable objects from an input catalog containing more than 100,000 entries while taking into account the object's sky visibility at each telescope's position.
Typically, each observation is first reviewed by the astronomer prior to being processed.
The review process involves examining nearly 20 GB of raw image data per telescope to detect and confirm the presence of transient astronomical events that are of interest to astronomers.
To free astronomers from these phases with tasks, we developed the StarWhisper Telescope (SWT) system, a LLM~\cite{ouyang2022traininglanguagemodelsfollow, InstructGPT} agent-based astronomical observation system designed to assist astronomers by efficiently managing observations across a telescope array and saving astronomers' time.

While LLMs have demonstrated extraordinary capabilities in many fields, traditional LLMs, such as Qwen~\cite{bai2023qwen,yang2024qwen2technicalreport}, the GPT series~\cite{openai2024gpt4technicalreport}, and DeepSeek~\cite{deepseekv2}, have not been equipped with tools or trained with specific data for astronomical observations, and more likely to present well-known celestial objects, no matter it can be observed or not (See Supplementary Section 1, Observation planning by Qwen).
The SWT system addresses these challenges by integrating function calls to expand the capacities of LLMs and LLM-enhanced workflows to infuse domain expertise.

Function calls enable LLMs to interact with external tools or systems by recognizing predefined operations, extending their capabilities beyond text generation to perform tasks or retrieve real-time information. This functionality is developed through training on datasets pairing input instructions with expected outputs, enhancing the models' instruction-following abilities~\cite{yang2024qwen2technicalreport,openai2024gpt4technicalreport}. When equipped with function call capabilities, LLMs are often referred to as LLM-based agents, or simply agents in these paper~\cite{andreas2022agent,qin2023toolllmfacilitatinglargelanguage,wang2024mintevaluatingllmsmultiturn,sinha2024LLMeval,langchain_chatchat,qian2024experientialcolearningsoftwaredevelopingagents}.

LLM-enhanced workflow connects steps in a series or parallel configuration, incorporating tools, Application Programming Interfaces (APIs), and LLMs to handle complex, multi-stage tasks and do function calls. Compared to traditional workflows~\cite{n8nref}, it can process unstructured data, make flexible judgments, and adaptively link each step of the process.  This approach increases the credibility and usability of the system through the operation and interaction logic of natural language, while maintaining robust input/output structures~\cite{wei2023CoT, langchain_chatchat, andreas2022agent}. The workflow in this manuscript infers the LLM-enhanced workflow.

\begin{figure*}
    \centering
    \includegraphics[width=0.95\linewidth]{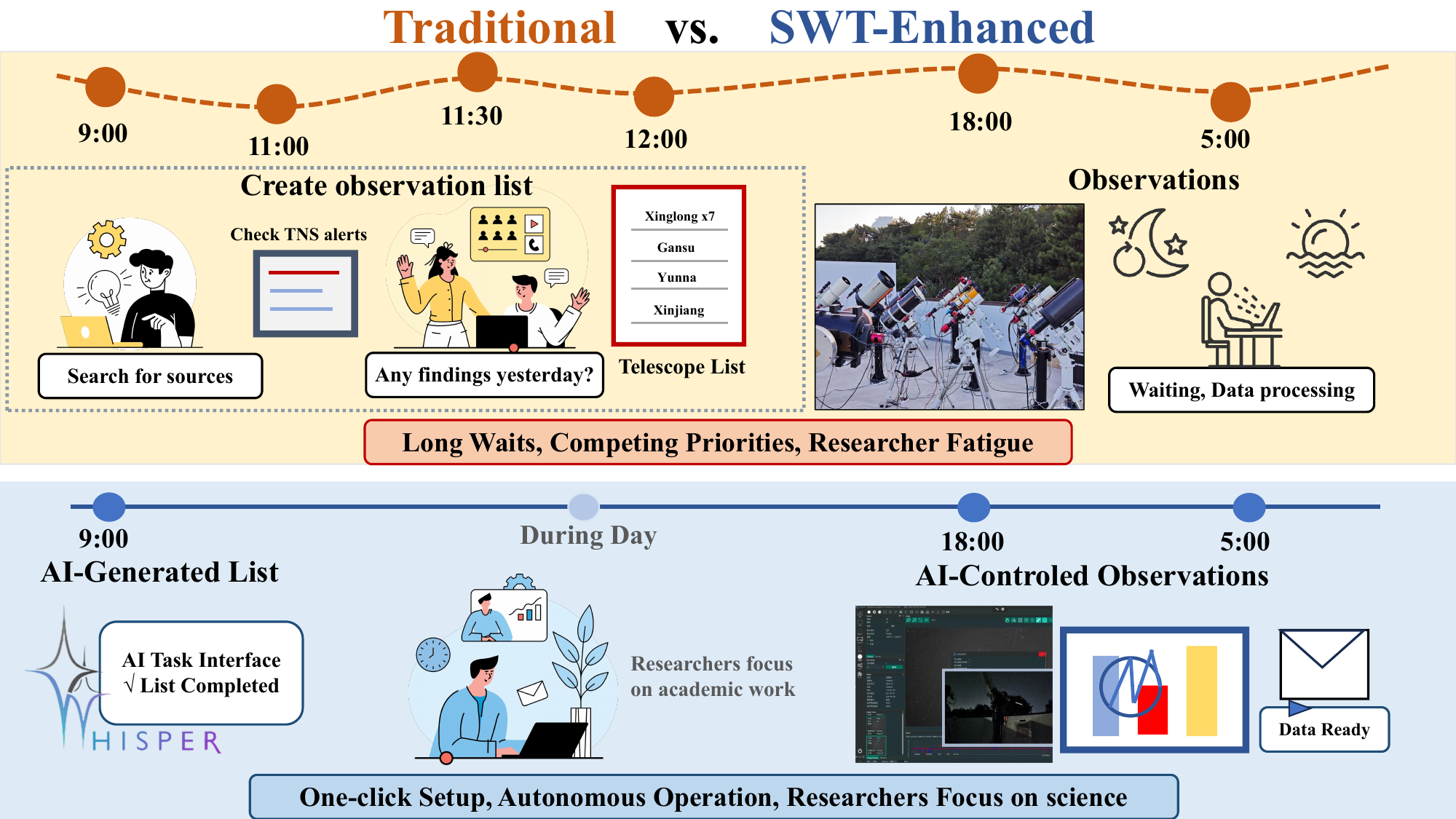}
    \caption{\textbf{Workflow comparison between traditional observation methods and the SWT enhanced system.} The TNS stands for Transient Name Server. }
    \label{fig:OB_chouxiang}
\end{figure*}

The SWT system is applied to the Nearby Galaxy Supernovae Survey (NGSS, Section~\ref{sec:NGSS}) to help the observers save time (Figure~\ref{fig:OB_chouxiang}). The NGSS project is a time-domain survey with 10 amateur-level telescopes, with a simplified telescope organizational structure similar to GOTTA.  In summary, the SWT observation module (Section~\ref{sec:Mainagent}) consists of observation planning(Section~\ref{sec:OBlist}), observation control (Section~\ref{sec:OBcontrol}) and a data pipeline (Section~\ref{sec:Data_pipeline}). Before and after observations, SWT provides recommendations to observers through automated messages (Section~\ref{sec:AI_suggest}). We show the observation and execution of the SWT system in Section~\ref{sec:result}, and discuss its strengths and weaknesses in Section~\ref{sec:pros}. The solutions to the weaknesses are discussed in Section~\ref{sec:futureplan}. The related works are referred in Supplementary Section 2, Related Works.

%We also discuss experiments conducted with the GOTTA pathfinder telescope (mST, \cite{huang2025minisitianarraypathfindersitian,Han2024mST,He2024mSTOB,Wang2024mSTOCS}) in Section~\ref{sec:mst}, demonstrating the necessity of employing agent workflows in telescope arrays to reduce human effort. Finally, we outline our future plans for implementing a fully integrated agent workflow in observations in Section~\ref{sec:futureplan}.

\section{Nearby Galaxy Supernovae Survey} \label{sec:NGSS}

As outlined in Section~\ref{sec:intro}, we have developed an agent-based system named SWT, aimed at easing the challenges associated with astronomical observations, particularly focusing on large-scale time-domain surveys such as the GOTTA (More information are provided in Supplementary Section 3, GOTTA). Although GOTTA is still under development, we test the SWT system's performance through a similar survey called the NGSS, which features multiple sites and telescopes, thereby serving as an ideal testing ground for GOTTA.

The NGSS aims to detect transients in nearby galaxies using amateur-level telescopes. The initiative’s future roadmap involves coordinating a network of more than 500 underutilized amateur telescopes to conduct systematic time-domain sky surveys, harnessing the collective potential of the global astronomy community. Similar to the GOTTA, its primary function is to identify transient astronomical events, but on a smaller scale, suitable for amateur equipment.

Several amateur telescopes have been installed at the Xinglong Observatory in Hebei Province ($40.393^\circ$ N, $117.575^\circ$ E), with additional telescopes located in Xinjiang ($43.522^\circ$ N, $88.577^\circ$ E), Gansu ($35.678^\circ$ N, $106.848^\circ$ E), and Yunnan($23.914^\circ$ N, $102.653^\circ$ E) provinces in China. These telescopes, detailed in Table~\ref{tab:NGSS}, are equipped with monochrome CMOS sensors and LRGB photometric filters. With their relatively large Field of View (FoV) and apertures under 250mm, they are particularly suited for detecting bright supernovae in nearby galaxies. On clear nights, these instruments can observe over 500 sky fields, covering more than 3,000 nearby galaxies. Larger amateur telescopes with apertures exceeding 250mm and narrower FoVs are used primarily for follow-up photometry, complemented by professional telescopes like the 216-cm telescope at Xinglong Observatory for follow-up spectroscopy and precise photometric observations. Considering the power of these telescopes, this work focuses on galaxies within 50 Mpc \cite{david24_50mpcgal} and associated transient phenomena.

%To optimize the use of each telescope across different sites, tailored observing lists and programs are essential. For each survey telescope, 2-minute exposures in the L band are employed for survey observations, ensuring non-overlapping observation lists to maximize source coverage. Observing priorities vary by location; for example, telescopes in Yunnan focus on sources with lower declinations. This strategic planning allows for the efficient use of existing telescopes and facilitates rapid integration of new telescopes into the network.

To reflect the operational preferences of the amateur astronomy
community as much as possible, NGSS operations utilize Nighttime Imaging 'N' Astronomy (N.I.N.A.) software, a widely-used automation tool among amateur astronomers. This software integrates the control processes, allowing the observer to connect all equipment, such as the camera, telescope, dome, and filter wheel, to execute the observation from a given list in a specific format (a ninaTargetSet file). N.I.N.A. allows users to develop custom plugins to meet specific needs, for example, the three point polar alignment. The software requires a list file to execute observations and uses ASCOM drivers to control the entire telescope system, performing fully automatic operations including slewing, auto-focusing, astrometric solving, capturing images, switching targets, and returning to park position upon completion.

\begin{table}[ht]
  \centering
  \begin{tabular}{ccccc}
    \hline\noalign{\smallskip}
    \begin{tabular}{c}
      Telescope \\
      Name
    \end{tabular}        &
    \begin{tabular}{c}
      Field-of-View \\(deg$^2$)
    \end{tabular} &
    \begin{tabular}{c}
      Focal Length \\
      (mm)
    \end{tabular}        &
    \begin{tabular}{c}
      Pixel Scale \\
      (arcsec $\rm pixel^{-1}$)
    \end{tabular}        &
    \begin{tabular}{c}
      Telescope \\
      Location
    \end{tabular}                                                        \\
    \hline\noalign{\smallskip}
    xl-106                    & 10.14 & 530  & 1.46  & Xinglong   \\
    xl-130                    & 2.91  & 990  & 0.784 & Xinglong   \\
    xl-130-2                  & 3.43  & 909  & 0.853 & Xinglong   \\
    xl-180                    & 4.80  & 502  & 1.54  & Xinglong  \\
    xl-203                    & 1.18  & 864  & 1.29  & Xinglong  \\
    xl-250                    & 0.18  & 2575 & 0.301 & Xinglong   \\
    xl-c14                    & 0.16  & 2717 & 1.14  & Xinglong \\
    gs-150                    & 0.83  & 678  & 0.73  & Gansu     \\
    yn-90                     & 2.27  & 601  & 1.85  & Yunnan     \\
    wlmq-107                  & 2.47  & 700  & 2.8   & Xinjiang    \\

    \noalign{\smallskip}\hline
  \end{tabular}
  \caption{\textbf{Details of Amateur Telescopes Used for NGSS.} The telescope name contains two parts, 'xl', 'yn', 'gs', 'wlmq' is the place, and the rest shows the diameter or the model number of the telescope. }\label{tab:NGSS}
\end{table}
% gkd-700 & 0.12 & 4561 & 0.61 & Beijing & Follow-up \\
\section{Methods} \label{sec:Mainagent}

LLM-based agents relieve researchers of repetitive tasks in scientific workflows, enabling a strategic shift toward high-impact intellectual contributions. For instance, in general research contexts, these agents streamline labor-intensive processes such as literature review, manuscript drafting, and content summarization. In astronomy, specialized agents further augment productivity by automating critical but there are time-consuming responsibilities, including preparing observation proposals, designing and executing observational strategies.

The SWT system is designed to orchestrate and interface with the NGSS telescope network through AI-driven agents, enabling autonomous control and adaptive observation workflows. Several functional workflows are interconnected with workflows and API-driven communication protocols.

\begin{figure*}
    \centering
    \includegraphics[width=0.85\linewidth]{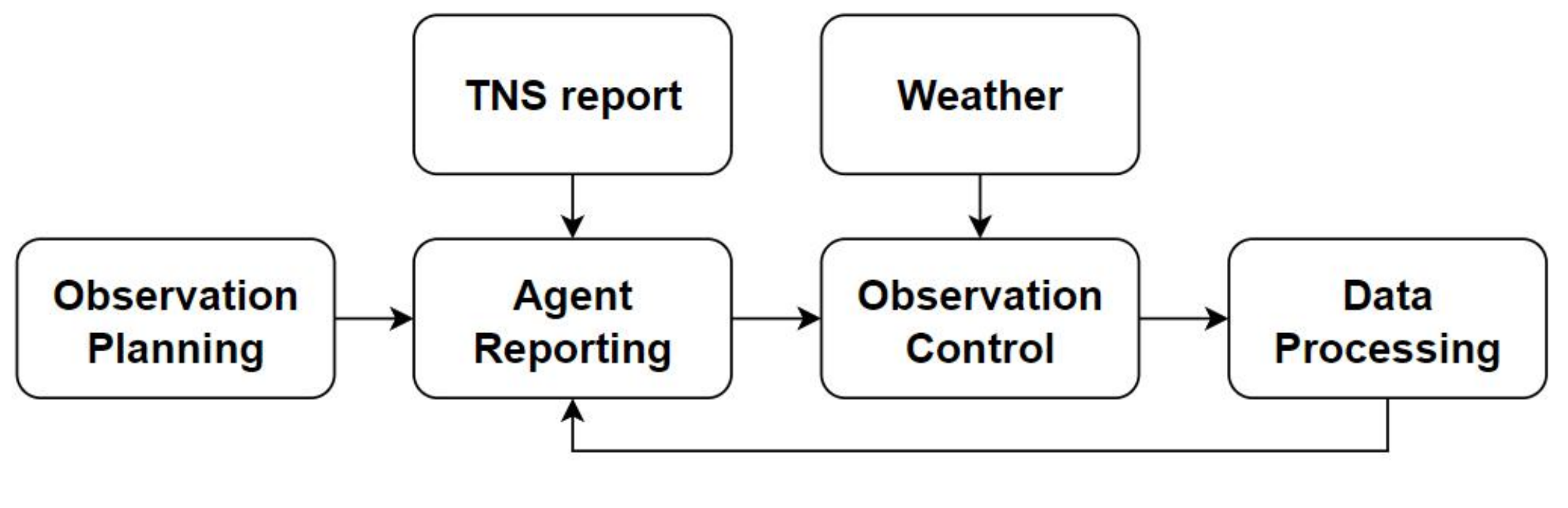}
    \caption{\textbf{The main structure of the SWT system.} It begins with Observation Planning, where initial observation lists are generated. The process then moves to Agent Reporting, which incorporates updates from the Transient Name Server (TNS) report and the transients found by the data processing module. This procedure also allows manual revision and injects new targets into the observation list.  These plans are then operated by the telescopes via the Observation Control module, in which the weather station can send commands to halt the observation. The Data Processing phase is designed to find transients from a data pipeline. }
    \label{fig:OB_NGSS_main}
\end{figure*}

In this section, we present the architecture of the SWT system, which comprises a central agent (Section \ref{sec:prompt-engineering}) and four sequentially connected workflows (Figure \ref{fig:OB_NGSS_main}): observation planning (Section \ref{sec:OBlist}), telescope control (Section~\ref{sec:OBcontrol}), data processing (Section \ref{sec:Data_pipeline}) and agent reporting (Section \ref{sec:AI_suggest}). The central agent is integrated with all workflows, providing comprehensive control over each component.

Regarding the LLMs used in the SWT system, agents and workflows, we utilize Qwen-2.5 ~\cite{qwen2.5} as our base model. The details of the architectures are provided in the Supplementary Section 4, Detail and Prompt Design of NGSS Agent; Section 5, Observation Planning and Section 6, Auxiliary tools.

\subsection{Central Agent} 
\label{sec:prompt-engineering}

The SWT system adopts a modular workflow architecture designed to automate astronomical observation tasks while maintaining flexibility for human intervention. Built upon a distributed processing paradigm, the SWT system integrates 8 specialized workflows that operate through coordinated API communications, parameter-driven task routing, and AI-enhanced decision layers. 

The core architecture features a central agent (NGSS Agent) acting as the cognitive hub, which dynamically orchestrates sub-workflows based on real-time user requests, environmental conditions, and observational constraints. 

The eight workflows and their structure are as follows:

\begin{itemize}
    \item \textbf{Central Workflow}: The central workflow orchestrates the overall system operations, including setting parameters, managing memory buffers, handling LLM models, and invoking tools via a central Agent.
    \item \textbf{Observation Planning}: This workflow generates observation plans, communicates with the local server, and produces a log Uniform Resources Locator (URL) and the task Universally Unique Identifier.
    \item \textbf{Observation List Query}: This workflow allows users to query observation plans for a specific site and date.
    \item \textbf{Transient Loading}: This workflow handles the viewing and addition of transient sources and provides links to identification and image results.
    \item \textbf{Target Addition}: This workflow enables users to add new astronomical targets (including targets from Transient Name Server, TNS) to the observation list and provides a summary of the results.
    \item \textbf{Plan Loading}: This workflow loads observation plan files into the telescope's N.I.N.A. software.
    \item \textbf{Telescope Control}: This workflow controls the start and end of telescope observation actions via API calls.
    \item \textbf{Weather Monitoring}: This workflow retrieves and summarizes weather data to ensure suitable observing conditions.
\end{itemize}

We show the details and the prompt design for the main agent in Supplementary Section 4, Detail and Prompt Design of NGSS Agent.

\subsection{Observation Planning}
\label{sec:OBlist}

\begin{figure*}
    \centering
    \includegraphics[width=0.85\linewidth]{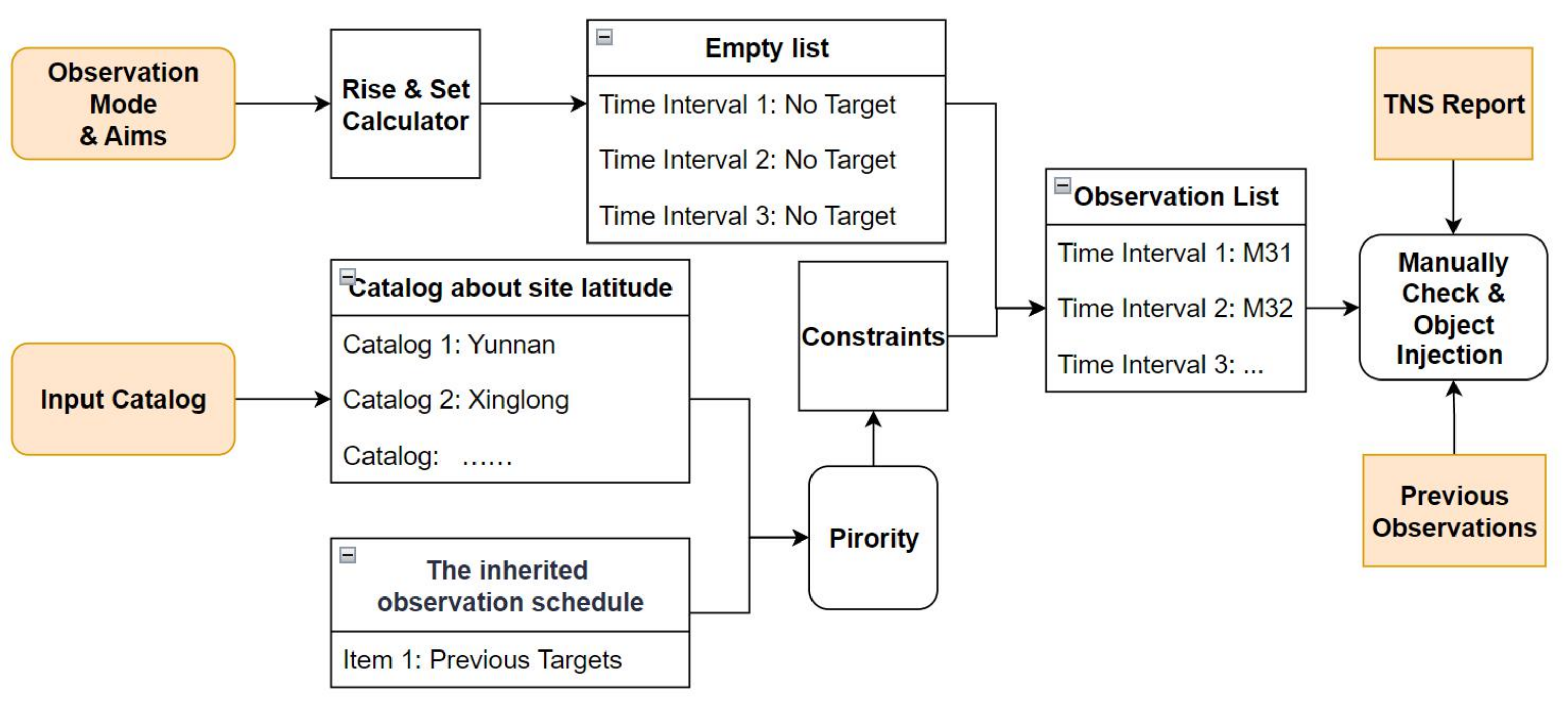}
    \caption{\textbf{Schematic diagram illustrating the construction of the observation list for the NGSS.} Regular rectangles denote tools, while rounded rectangles represent agents powered by LLM. All orange boxes indicate input information. The constraints tool calculates both altitude and moon distance limitations. Once objects are manually added, the finalized list is saved to the server.}
    \label{fig:OB_NGSS}
\end{figure*}

The observation planning workflow is designed to create observation lists for 10 telescopes across four sites in NGSS. Further details of this workflow can be found in Figure \ref{fig:OB_NGSS}.

The observation planning is done by a function call to trigger the workflow, and sequentially trigger the generation of a day's plan. The workflow first generates an empty observation list based on the configuration files. The configuration file includes the pre-set observation parameters, including but not least the exposure, tracking, focusing and guiding settings, filter configurations, slewing and readout time, is designed to improve the workflow's robustness. The workflow then fills the empty observation lists with sources in the input catalog, or the previous observation lists. The site position and several constraints will be considered during planning, and the final plan will be revised by the astronomer.

More details of Observation Planning are shown in Supplementary Section 5, Observation Planning.

\subsection{Observation control}
\label{sec:OBcontrol}
\begin{figure*}
    \centering
    \includegraphics[width=0.85\linewidth]{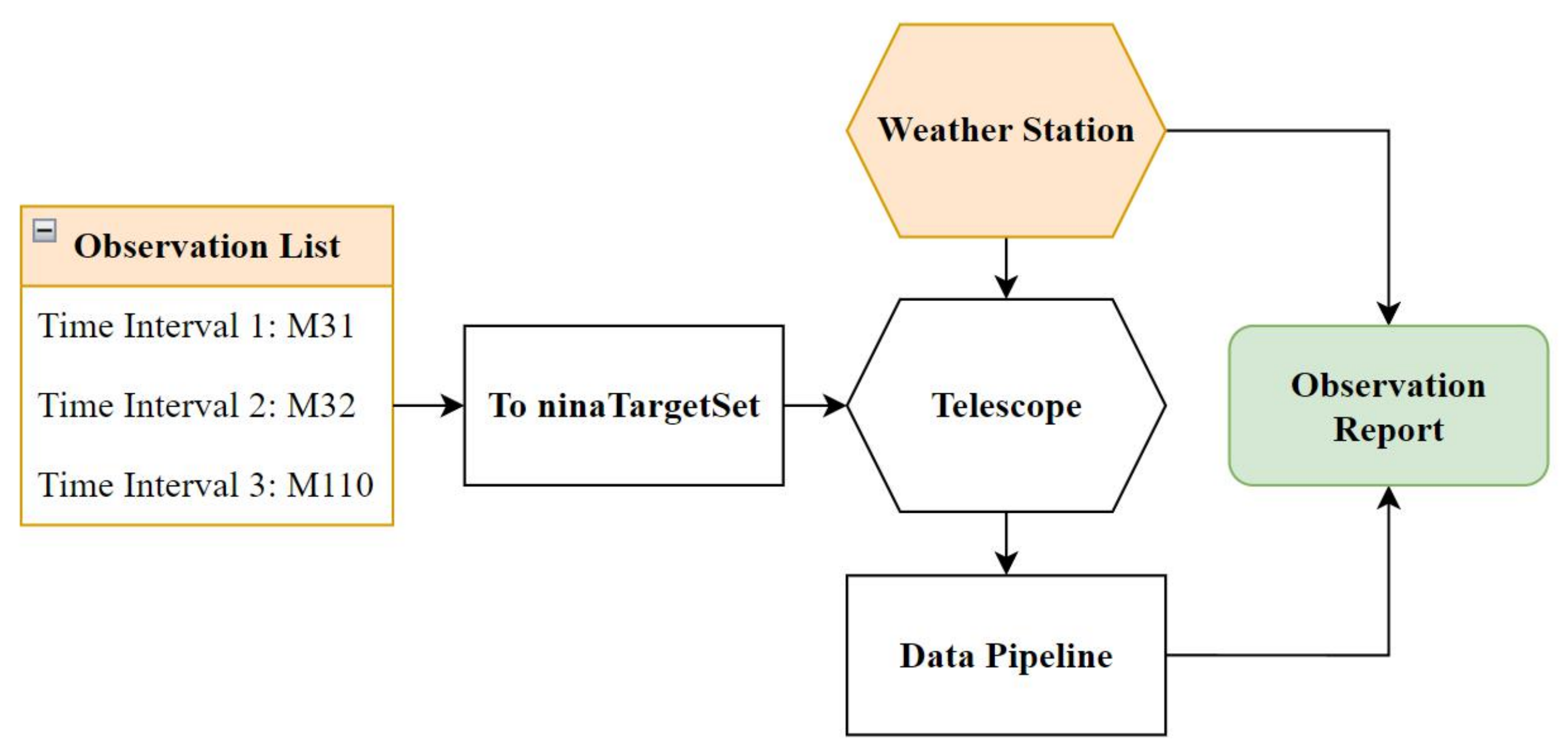}
    \caption{\textbf{Schematic diagram of the observation component in the NGSS.} Regular rectangles denote tools, while hexagons represent hardware facilities. All orange boxes indicate input information, and green boxes correspond to output results. The observation list will be transformed into a ninaTargetSet file for N.I.N.A. to read and take control of the observation. The data will be simultaneously processed by a data pipeline. The weather station at Xinglong Observatory can automatically halt the observation by commands to N.I.N.A.. The transients found will be reported after data processing.}
    \label{fig:OB_NGSS2}
\end{figure*}

After the generated observation list is reviewed by an astronomer, it will be transferred to the appropriate format and sent to N.I.N.A. to start the observation (Figure \ref{fig:OB_NGSS2}).

We built a N.I.N.A. plugin based on the Site Plugin~\cite{siteplugin}(\url{https://nighttime-imaging.eu/docs/master/site/contributing/plugins/}) with the functions of loading ninaTargetSet, an observation list format that can be read by N.I.N.A., and starting and stopping observation via a User Datagram Protocol message. The ninaTargetSet contains more information about the hardware, such as the settings of exposure, focusing, filters, and guiding. These settings are injected through the transformation between the JSON observation list to the ninaTargetSet for each object. By default, every two hours, the automatic focusing will be enabled, and 30 new bias images will be taken.

At the predetermined time (the end of astronomical twilight), the agent issue commands by sending corresponding messages, which then triggers the plugin to initiate observations. Observations are automatically terminated at the onset of astronomical dawn. Additionally, the API from the compact weather site is integrated with the agent, enabling the automatic suspension of observations in response to adverse weather conditions.

\subsection{Data pipeline}
\label{sec:Data_pipeline}

The raw image taken by the telescopes will be processed in the data pipeline to find transients and send alerts. Usually, the follow-up spectroscopy observation proposal will be sent to the 216-cm telescope (at Xinglong Observatory), and manually do the follow-up observation since its not linked to the SWT system. 

Xinglong - Observatory Popular Science Telescope Pipeline (X - OPSTEP, \cite{xopstepgit}) is the data pipeline for NGSS. It can automatically and real-time process the raw image data, including image pre-processing, astrometry, photometry, and image subtraction.

After the observation process by N.I.N.A., the raw image will be sent back to the server at the Xinglong observatory, then X-OPSTEP will be triggered as long as the new raw data is received. X-OPSTEP first performs bias and flat correction, and then solves for the image world coordinate system information with \texttt{Astrometry.net}~\cite{Lang_2010_astrometrynet}. After that, the photometry process will be done by SExtractor~\cite{bertin1996sex}, and the resulting catalog will be cross-matched with GAIA DR3~\cite{Gaiadr3vizier:I/355} to conduct flux calibration. The pipeline will then combine the same target for the same band By using SWarp~\cite{bertin2002terapix} with 3-$\sigma$ clipping median combination to get rid of contamination from any satellites. Finally, we use \texttt{HOTPANTS}~\cite{becker2015hotpants} to subtract the template image from the science image and cut out the target galaxies. See figure~\ref{fig:2025pk} for an example. A real-bogus model is applied to check the subtracted image. The model, similar to Shi et al., 2025\cite{Shi_2025SiTian}, using a $64 \times 64$ difference image as input, predicting through a ResNet + attention architecture, achieves 99.12\% validation accuracy.

\subsection{Suggestions Given by the AI agent}
\label{sec:AI_suggest}

The photometric data and image subtraction result will be read by the SWT system. The SWT system evaluates multiple options to assign, remove, or replace specific galaxies in the observation lists. Typically, it recommends that the user adopt the observation lists from the previous day to form continuous monitoring of these galaxies. These galaxies are flagged and reassigned to the observation list. Similarly, galaxies with detected transients are also highlighted and included in the subsequent observation list for continuous imaging.

Once a transient is detected, its coordinates and associated information are transmitted to the user via a web interface. If the Agent is unable to retrieve any information about the object, it will recommend that the user submit the object to the TNS. The SWT system will also send a message about the transient to the XingLong chat group, including information about the source and an observation suggestion generated by the LLM.

We show some auxiliary tools in Supplementary Section 6.

\section{Results} \label{sec:result}
The SWT system has been actively operating on NGSS telescopes since October 2024, successfully detecting several transients. The majority of these transients are initially reported by other open surveys, and we cross-matched them with our observational data, and found some transients bright enough to be detected. Some transients have observing time earlier than the discovery time reported, but not detected due to large background noises or detected before explosion. Details of these transients are shown in Table~\ref{tab:transient}.

\begin{table}[ht]
    \centering
    \begin{tabular}{cccccc}
        \hline\noalign{\smallskip}
        \begin{tabular}{c}
            Transient \\
            Name
        \end{tabular}      &
        \begin{tabular}{c}
            Our \\First \\Detection
        \end{tabular} &
        \begin{tabular}{c}
            TNS \\Report \\Date
        \end{tabular}     &
        \begin{tabular}{c}
            Transient \\Type
        \end{tabular}      &
        \begin{tabular}{c}
            Redshift \\( z )
        \end{tabular}      &
        \begin{tabular}{c}
            Discovery \\
            Telescope
        \end{tabular}                                                                      \\

        \hline\noalign{\smallskip}
        SN2024xin               & 2024-10-07 & 2024-10-05 & SN Ia           & 0.019 & xl-106   \\
        SN2024xlh               & 2024-10-07 & 2024-10-06 & SN II           & 0.03  & xl-106   \\
        SN2024xli               & 2024-10-07 & 2024-10-06 & SN Ia           & 0.037 & xl-106   \\
        SN2024xqe               & 2024-10-10 & 2024-10-09 & SN Ia           & 0.034 & xl-106   \\
        SN2024xvg               & 2024-10-10 & 2024-10-10 & SN Ia           & 0.03  & xl-250   \\
        AT2024abqt              & 2024-11-21 & 2024-11-21 & CV              & -     & xl-130-2 \\
        SN2024advj              & 2024-12-11 & 2024-12-11 & SN IIn          & 0.017 & xl-106   \\
        SN2025bl                & 2025-01-04 & 2025-01-04 & SN II           & 0.009 & xl-130-2 \\
        AT2025pk                & 2025-01-17 & 2025-01-17 & Flare Candidate & -     & xl-130-2 \\

        \noalign{\smallskip}\hline
    \end{tabular}
    \caption{\textbf{Details of Transients Detected by NGSS.} The first detection date is the date on which the transient was detected based on the image subtraction. The '-' in redshift means no data reported. The TNS is the Transient Name Server plateform, and the types are Super Novae (SN), Cataclysmic Variable (CV).}
    \label{tab:transient}
\end{table}

As shown in Table~\ref{tab:transient}, during the early test observations, the actual detection time for transients lagged approximately one day behind the TNS reporting time. This delay is primarily attributed to the manual maintenance at the beginning of NGSS, at the time when the manpower can only afford two to three telescopes. 

However, after deploying SWT system, the detection time of brighter transients, such as AT2024abqt, SN2024advj, and SN2025bl, trailed the TNS reporting time by only a few hours. Notably, AT2025pk, a flare star, represents the first transient event successfully identified through the SWT system. 
 
\begin{figure*}
    \centering
    \includegraphics[width=0.85\linewidth]{figures/AT2025pk.pdf}
    \caption{\textbf{The discovery confirmation images of AT2025pk.} The images are shown from left to right are template, science, and difference image. The title contains object name (OBJNAME), background magnitude (BMAG), the pixel position of AT2025pk, Right Ascension (ra), Declination (dec), and the hour angles in hour-minute-second (hms) and degree-arcminute-arcsecond (dms).}
    \label{fig:2025pk}
\end{figure*}

Figure~\ref{fig:2025pk} shows the transient source identification images provided by the SWT system. Bright transient sources are evident in both the science and difference images. Through visual inspection, we search for suspicious bright spots in the images and cross-check to exclude asteroids or comets, thereby enabling the discovery of transients.

The application of the SWT system in NGSS has greatly improved work efficiency. The observation plan was completely generated manually in the beginning phase of NGSS. To quantitatively evaluate the performance of the SWT system, we conduct a controlled comparison between manually generated and SWT-generated observation plans under identical conditions. The observation site is considered as Xinglong. The results (Table~\ref{tab:manual_vs_SWT}) show that SWT reduce the planning time from about 1.5 hours (PhD level) to less than 1 minute, with better target coverage counts and zero conflicts, demonstrating both efficiency and robustness.

\begin{table}[h]
\centering
\label{tab:manual_vs_SWT}
\begin{tabular}{p{4cm}|p{4cm}|p{4cm}}
\hline
Metric & Manual (PhD student) & SWT System \\ 
\hline
Planning time per telescope & 1 - 1.5 hours & $<$1 minute \\ 
\hline
Number of galaxy coverage on the same day & 2000 - 2500 & 2500 - 3000 \\ 
\hline
Conflict rate per list & 1 - 3  & 0 \\ 
\hline
Requires manual adjustment & Frequently needed & Not required \\ 
\hline
Reproducibility / Consistency & Depends on operator & Fully deterministic \\ 
\hline
Adaptability to constraints & Limited & Sites, visibility, and moon phase \\ 
\hline
\end{tabular}
\caption{\textbf{Comparison between manual and SWT system for observation planning.} The manual plan is generated by a PhD student, with galaxy coverage based on the combined 10-telescope array. The conflict rate reflects the number of galaxies observed more than once in a single list. The time refers to the period when observable galaxies can be distributed across the telescopes. Manual adjustment of the SWT plan is only applied when target changes are needed. In manual planning, it is difficult for humans to account for all constraints simultaneously.}
\end{table}

We evaluated the success rate of SWT system by cycling through the following prompts and querying the agent round by round: (1) Please make an observation plan; (2) Please check today's log of making observation plan; (3) Please check the observation plan; (4) Please check the transients; (5) Please add them to the observation list; (6) Please load the observation lists to each telescope. These queries are commonly used among observers when interacting with the agent. The programs for starting or stopping observations have been prohibited from running during the test to avoid any potential risk of system damage.

The agent has been asked 7,620 rounds, of which 4,194 involved tool calling, and the tool has been called 2,962 times successfully, showing the overall function call success rate is approximately 70.5\%. The number of function calls is less than the number of queries because not all queries need to make function calls. The success rates of the observation planning and observation list querying tools have achieved 100\%. For loading transient or input sources and telescope-controlling tools, the success rates range between 60\% and 70\%. The loading observation plan for the telescope tool has a relatively low success rate of approximately 30\%. These issues are mainly caused by the network latency, and lead to an overtime failure. The test is conducted over two days, and the total running cost is about 871.6 minutes, involving about 58.6 million tokens cost. Tokens are a basic unit of text processed by LLMs, providing the context for usage in our workflow. This amount of tokens is about 14 dollars at cloud services.

The example of usage of the SWT is shown in Supplementary Section 7.

\section{Discussion}
\label{sec:dis}
In the previous Section \ref{sec:Mainagent}, we show the construction of the SWT system. In this section, we first introduce the strengths and challenges of the system in Section~\ref{sec:pros}. We then make a future plan at Section~\ref{sec:futureplan} to solve the technical issues we faced, and the path to an AI Astronomer.

\subsection{Strengths and Weakness}
\label{sec:pros}

Astronomical observations have traditionally demanded specialized expertise; however, our agent-integrated system substantially lowers the entry barrier by enabling natural language interaction and real-time process monitoring. This innovative approach not only accelerates the onboarding process for novice observers but also empowers experienced astronomers to efficiently manage multiple telescopes simultaneously, thereby considerably improving observational productivity.

The development of standardized protocols, such as Agent-to-Agent~\cite{a2a_project}, which allow agents to interact with another agent, among different agents, lays a solid foundation for cluster control designed based on multi-agent system games. It enables seamless customization for a wide range of survey projects and adaptability to various observational strategies, including dynamic updates to observation lists and real-time control.

We also identify several limitations of the current system and discuss how they may be addressed or what lessons have been learned.

\paragraph{Automated hardware}
The dome of NGSS is not automated and must be opened or closed manually, rendering automatic observation halting partly ineffective. At Xinglong Observatory, this issue is mitigated by sending warnings to the observers of the GOTTA pathfinder, which shares the same dome. However, for telescopes located at other sites — typically housed in amateur domes situated on villagers’ rooftops — dome operations cannot be controlled remotely. Dome management at these locations relies entirely on the cooperation of local residents who rent out their rooftop space. Similarly, flat exposure must currently be taken manually, due to the lack of automation. At present, we have created standard template flat images to ensure basic data quality in Gansu, Yunnan, and Xinjiang telescopes. Looking forward, integrating automated hardware will be a feasible solution for telescopes at Xinglong Observatory. However, for remote amateur sites, achieving full automation remains a considerable challenge due to logistical and infrastructural constraints.

\paragraph{Hardware failures}
The SWT system is currently unable to automatically resolve hardware failures. Common issues include focuser freeze due to cold weather, wiring disconnections, and computer memory crashes, with the latter two being the most frequent. At present, these problems are resolved manually by on-site observers. In the future, the system could be enhanced to provide troubleshooting suggestions through an intelligent agent. This would be achieved by analyzing observation logs written by observers, which the SWT system could access using Retrieval-Augmented Generation~\cite{lewis2021rag} or a Model Context Protocol. Retrieval-Augmented Generation is a technique that enhances LLM by retrieving information from a specified dataset, considerably improving accuracy and relevance in specific domains. Model Context Protocol is a protocol designed to define contextual interactions for tools or datasets, aiming to standardize communication and integration between models and external resources. For larger telescopes, continuous monitoring systems could be implemented to collect standardized telemetry data, allowing the agent to assess telescope status and detect anomalies in real time. However, for telescopes located outside of Xinglong Observatory — particularly those hosted in amateur setups — there are currently no feasible conditions for immediate or automated fixes due to limited infrastructure and remote locations.

\paragraph{Software crashes} 
The SWT system currently lacks the ability to automatically resolve low-level software crashes, such as those occurring in applications like N.I.N.A. and X-OPSTEP. These incidents are typically infrequent but often stem from human-driven overuse of server resources—such as running multiple memory-intensive programs simultaneously, which exceeds available storage capacity. Automated solutions for these types of failures are not considered a high priority, mainly due to their rarity and the challenges associated with changing operator workflows or upgrading server specifications. While agent-based monitoring of the software environment could potentially detect and mitigate such issues, it introduces considerable security risks by requiring privileged access (e.g., through tools like Code Interpreter). An alternative approach involves training a GUI agent \cite{hu2024guiagent} to monitor graphical interfaces and respond to anomalies. However, this method requires substantial GPU memory resources, making it impractical for deployment at scale in large observatory operations.

\paragraph{Standardized telescopes} 
The current lack of standardization among telescopes in the NGSS network introduces operational inconsistencies, particularly in manual observation practices. Two key challenges emerge for the SWT. First, the varying FoVs across different telescopes affect overall sky coverage. The input catalog is currently designed based on the smallest FoV to maintain uniformity, which results in larger FoV telescopes potentially observing adjacent or overlapping regions and inadvertently covering galaxies already targeted by smaller FoV instruments. This leads to a reduction in total effective sky coverage compared to manually generated plans. A new tool is being developed to shift the planning focus from individual galaxy selection to maximizing sky coverage, which is expected to improve this limitation. Second, differences in hardware reliability across telescope types result in varied failure patterns, making it more difficult to implement consistent fault detection and recovery strategies. Standardization of both hardware and software components is therefore essential not only for improving system usability and agent-instrument interaction, but also for enabling scalable deployment. As the NGSS network expands with more standardized nodes, it will achieve broader sky coverage, enhancing its scientific and educational capabilities. Furthermore, standardization reduces operational complexity by minimizing labor-intensive maintenance and improves data consistency by reducing instrumental variation across observational outputs.

\subsection{Future works} \label{sec:futureplan}

\paragraph{Edge Computing}

Integrating edge computing modules with telescope systems offers a promising solution to latency issues in server-to-server communication, enabling telescopes to make autonomous decisions and improving operational responsiveness and control stability. By deploying lightweight agent services and AI models locally—such as on industrial personal computers equipped with edge devices like the NVIDIA Jetson—real-time decision-making and task execution can be achieved with minimal reliance on remote servers or cloud infrastructure.

\paragraph{Observation monitoring}

A future observation monitoring module is crucial for ensuring the reliability of unattended observatories. It will use sensors to collect real-time data on the telescope’s physical state, including vibration (e.g., sounds from the equatorial mount), temperature, humidity, and electronic signals, to assess its operational status. When an anomaly is detected, the module will evaluate its severity and respond accordingly—automatically resolving low-severity issues through control commands or warning the observers with guidance for more serious problems.

We will collect operational data from the GOTTA prototype, a 1-m diameter prototype telescope for GOTTA (See more information in Supplementary Section 3, GOTTA), to generate training samples for various types of errors and malfunctions. These data will be used to construct a knowledge graph based on observer handbooks, aiming to enable automated monitoring and maintenance instructions—capabilities that are currently limited in existing systems. This effort will also include building a comprehensive dataset containing videos from cameras near the telescope, observation logs, and hardware manuals, which will serve as valuable training material for LLM-based agents assisting in system maintenance.

This dataset serves as a Retrieval-Augmented Generation-based memory module, integrating operational logs, interaction histories with observation assistants, and user behavioral profiles. These multi-source data are transformed into vector embeddings, enabling the retrieval of contextually relevant information during inference. By continuously learning from past successes, failures, and user-specific patterns, the system dynamically refines its responses, adapts to individual preferences, and supports distributed intelligence across edge computing nodes (e.g., telescope-mounted devices).

\paragraph{Path to AI Astronomer}

\begin{figure*}
    \centering
    \includegraphics[width=0.9\linewidth]{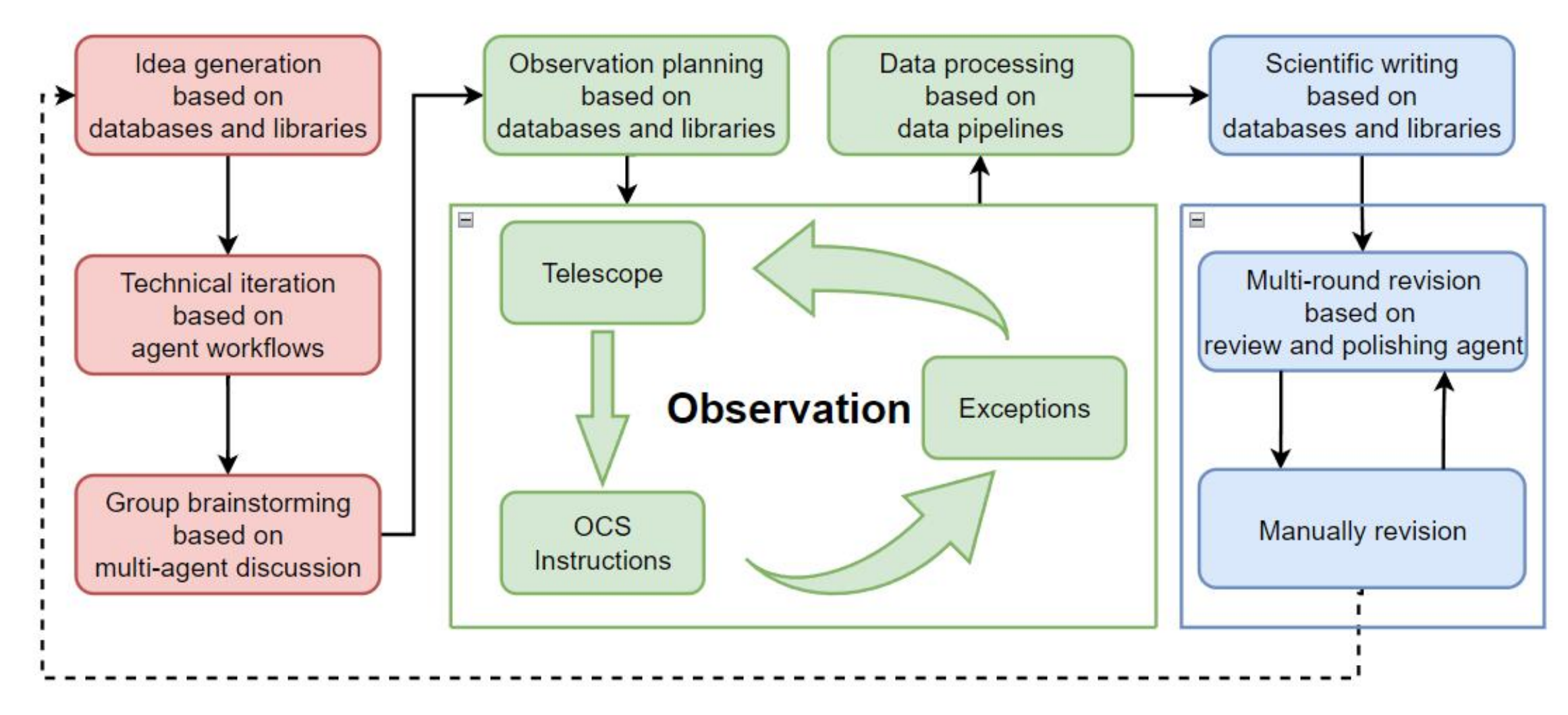}
    \caption{\textbf{The iterated process of AI Astronomer.} The red blocks are about the StarWhisper automatic scientific discovery system. The green blocks are the observation part, where we are trying to build the embodied telescope. The blue blocks show the StarWhisper scientific writing workflow. The OCS stands for the Observation Control System.}
    \label{fig:AIAstro}
\end{figure*}

Figure~\ref{fig:AIAstro} shows the methodology used by the AI Astronomer. For the telescopes that conduct sky surveys, it will get data from their real-time observation, and command the follow-up spectroscopy telescope to observe.

In Figure~\ref{fig:AIAstro}, the red blocks represent AstroInsight, an idea generation system which built on the integration of LLM-based agents and astronomical datasets. It will generate scientific ideas based on knowledge from databases such as ArXiv or Astrophysics Data System, and selects optimal methodologies for validation. 

For feasible plans requiring new data, the green blocks—enhanced by reasoning models like DeepSeek R1~\cite{deepseekr1}—support operations across the NGSS project, GOTTA, and 216-cm-telescope. This integration enables reasoned task dispatch and observation coordination, with acquired data feeding directly into AutoASTRO, the automated data processing and machine learning pipeline. The blue blocks represent the StarWhisper's scientific writing workflow, which transforms analytical outputs into structured narratives. This workflow forms a closed-loop feedback system, where the results of scientific writing are fed back into the idea generation module (red blocks), enabling continuous iteration and knowledge refinement.

\section*{Acknowledgments}
The research presented in this paper was generously funded by the National Programs on Key Research and Development Project, with specific contributions from grant numbers 2019YFA0405504 and 2019YFA0405000.  This research is supported by National Key R\&D Program of China (grant No. 2023YFA1609700 and 2023YFA1608304). Additional support came from the National Natural Science Foundation of China (NSFC) under grants NSFC-11988101, 11973054, 12090040, 12090041, 12403022 11903054, 12261141690 and 11933004. We also received backing from the Strategic Priority Program of the Chinese Academy of Sciences, granted under XDB41000000, XDB0550000, XDB0550100 and XDB0550102. Special acknowledgment goes to the China Manned Space Project for their science research grant, denoted by NO.CMS-CSST-2021-B07. The research is also supported by National Astronomical Observatories Chinese Academy of Sciences No. E4TG2001 and Key R\&D Program of Zhejiang (2024SSYS0006).

JFL extends gratitude for the support received from the New Cornerstone Science Foundation, particularly via the NewCornerstone Investigator Program, and the honor of the XPLORER PRIZE.

\section*{Author Contribution}
Cunshi Wang contributed the coding, agent construction, and the writing of the manuscript. 
Yu Zhang and Xinjie Hu contributed the coding, agent construction.
Yiming Mao contributed the test on NGSS project.
Xunhao Chen contributed the X-OPSTEP pipeline.
Pengliang Du contributed the data transmission and connections between telescopes to agents.
Rui Wang contributed setup and operation of the telescopes.
Yuyang Li contributed the design of the structure, prompt engineering for the SWT system.
Ying Wu contributed the design NGSS array and the design of N.I.N.A. plugin.
Hang Yang contributed the edge calculation of the system.
Yansong Li contributed the writing of the manuscript.
Beichuan Wang contributed the TNS search module and the observation of these telescopes.
Haiyang Mu, Xiaohan Chen, Shunxuan He, Hao Mo, Liyue Zhang, Lin Du, and Yunning Zhao contributed the operation and maintenance of SWT system on NGSS.
Jianfeng Tian, Liang Ge, Shengming Li, Xiaomeng Lu contributed the design of the structure on embodied telescope and the future of SWT system.
Zheng Wang, Jinhang Zou contributed the design of GOTTA pathfinder OCS and provide insights in the observation based on agents.
Yongna Mao, Yang Huang, Ningchen Sun, Jie Zheng, Min He, Yu Bai, Junjie Jin, Hong Wu, Jifeng Liu contributed the support the SWT system and administrative and resource support.

\section*{Data availability}
We have uploaded the codes, prompts, tools, and supporting materials to GitHub at \url{https://github.com/Yu-Yang-Li/StarWhisper/tree/main/NGSS}.

The StarWhisper's scientific writing workflow can be found in the Coze store at \url{https://www.coze.cn/store/agent/7383556199616135179?utm_source=ai-bot.cn}.  See \url{https://www.coze.cn/store/agent/7383556199616135179?utm_source=ai-bot.cn&bid=6f7h1ou4k7g19&post_id=7396577135529394202} for the user handbook.

\section*{Code availability}
See \url{https://github.com/Yu-Yang-Li/StarWhisper/tree/main/NGSS} for the usage of the codes. %The connection part codes should be configured by yourself. 

\section*{Ethics Declarations}
We declare no competing interests.

\section{Supplementary: Observation planning by Qwen}
\label{sec:comp_qwen}

The observation plans generated by popular use models often focus on well-known celestial objects, such as the Andromeda Galaxy (M31) or the Triangulum Galaxy (M33), without considering crucial factors like the season or the observation time specified in the prompt.

We asked Qwen-max-202412 model, the model with the highest ability for Qwen series models at December 2024, to show us 10 galaxies to observe in Beijing, at 16th Jan. 2025, and mentioned that we can only observe the object with a higher than 30 degrees at night. The result is shown in Table \ref{tab:Qwen_test}. 

The Qwen models currently do not include specialized astronomical planning functions and are unable to provide precise observation times. Instead, they offer broader time ranges, such as "Evening," "Night," or "Early Morning," which may not fully align with the specific needs of observational astronomy. Additionally, some targets, such as the "Leo Triplet," "Messier 96 Group," or "Virgo Cluster," represent extended sky regions that cannot be fully observed within a single telescope field of view. Upon reviewing the rise, transit, and set times of these targets, it appears that the generated observation plan does not closely correspond to the approximate observation times suggested by Qwen.

\begin{table*}[htbp]
    \centering
    \begin{adjustbox}{angle=90}
    \begin{tabular}{ccccc}
        \toprule
        Approximate time  & Target name & Raise time & Decent time & Transit time\\
        \midrule
        Evening & Andromeda Galaxy (M31) &  -  &  $\sim 22:40$   & $\sim 17:00$ \\
        Evening & Cigar Galaxy (M82) &  $\sim 18:40$  &  -  & $\sim 02:00$ \\
        Evening &  Bode's Galaxy (M81) &  $\sim 19:00$  &  -  & $\sim 02:00$  \\
        Night &  Pinwheel Galaxy (M101) &  $\sim 00:20$  &  -  & $\sim 06:40$  \\
        Night &  Owl Nebula (M97) & $\sim 21:40$  &  -  & $\sim 04:00$   \\
        Night &  Leo Triplet (M65, M66, NGC 3628) & $\sim 23:40$  &  -  & $\sim 04:00$   \\
        Night &  Messier 96 Group (including M95, M96, and M105) & $\sim 21:10$  &  -  & $\sim 03:00$   \\
        Night &  Sombrero Galaxy (M104) & $\sim 03:00$  &  -  & $\sim 05:00$  \\
        Night &  Coma Berenices Cluster (Melotte 111) & $\sim 00:00$  &  -  & $\sim 05:00$ \\
        Early morning &  Virgo Cluster (including M84, M85, M86, M87, etc.) & $\sim 00:50$  &  -  & $\sim 05:00$\\
        Early morning &  Antennae Galaxies (NGC 4038 / NGC 4039) & $\sim 03:50$  & $\sim 05:20$  & $\sim 04:30$ \\
        Early morning &  Black Eye Galaxy (M64) & $\sim 01:00$  &  -  & $\sim 05:20$ \\
        \bottomrule
    \end{tabular}
    \end{adjustbox}
    \caption{\textbf{The observation list given by Qwen.} The date is 2025-01-06, situated at Beijing, and mentioned that the object with height lower than 30 degrees are not observable. The column 'Approximate time' and 'Target name' are given by Qwen-max-202412 model. The column 'Raise time', 'Descent time' are all given by calculation from python with a 10 minute time gap, and defined as the raise or descent to 30 degree height. For all the times, '-' means the time is not at night. For example, the raise or descent happens at noon. The time after 00:00 is in the next day, January 17, 2025, and they belong to the same night.}
    \label{tab:Qwen_test}
\end{table*}

\section{Supplementary: Related works}
\label{app:relatedworks}
Here, we summarize several related works:

\subsection{Large language models}
LLMs have demonstrated extraordinary capabilities in natural language conversations~\cite{openai2024gpt4technicalreport, bai2023qwen, deepseekv2, glm2024chatglm}, coding, and even contributing to scientific discoveries~\cite{Wang2023Scientificdiscovery}, especially through the use of chain-of-thought prompting~\cite{wei2023CoT}.
The GPT series models~\cite{InstructGPT,openai2024gpt4technicalreport} have been leaders among LLMs, excelling in various aspects such as chat models, instruction-following ability, coding skills, etc. They are also among the first models used to construct agents~\cite{langchain_chatchat}. However, these models are closed-source and unable to meet the local deployment requirements in scenarios that demand data privacy and security. As an alternative, we mainly used the open-source models as the LLM engine, especially the DeepSeek series models~\cite{deepseekv2} and the Qwen series models~\cite{bai2023qwen,yang2024qwen2technicalreport}.

%
%As an alternative, we use the DeepSeek series models~\cite{deepseekv2} and the Qwen series models~\cite{bai2023qwen,yang2024qwen2technicalreport}. The DeepSeek v2 model have higher ablility than Qwen 1.5 72B model~\cite{deepseekv2} and the Qwen 2 7B model are comparable~\cite{yang2024qwen2technicalreport} in evaluations, with no significant differences in performance; however, the cost per token is much cheaper by comparison\footnote{\url{deepseek.com}}. However, with the release of Qwen 2.5, from a practical experience standpoint, the Qwen 2.5 instruct 14B model significantly outperforms the DeepSeek series in instruction-following ability and shows a strong tendency to invoke tools.
%

In terms of agents, Langchain ChatChat~\cite{langchain_chatchat} is driven by evolving technical requirements and practical constraints, implementing basic agent interactions through function registration. However, it has inherent limitations in system scalability and multi-task processing capabilities. AgentScope~\cite{agentscope} is a code-driven multi-agent orchestration paradigm. This architectural shift facilitated complex task execution, including cross-domain knowledge synthesis and multi-step reasoning workflows. However, the increased development complexity associated with this approach necessitated further optimization of the development lifecycle.

Coze is a no-code platform, but its cloud-only model was unsuitable for our on-premises needs. Dify~\cite{dify} is a containerized solution deployable via Docker. Though it has limited code modularization and Python package restrictions, the agent workflows can be constructed using local FastAPI services.

n8n~\cite{n8nref} has strong local integration and extensible AI interface. Although its native AI features are limited, combining it with Dify enabled a hybrid architecture that balances efficiency and complexity. Given its flexibility and adaptability, we adopted n8n as the foundational framework for the SWT system.

\subsection{Astronomical observation}

The traditional means for observation execution is via the OCS interfacing with instruments. Over time, improvements are usually made to the control system by adding in more automation~\cite{nguyen2024simonsocs,koopman2024simonsocs}. The philosophy behind OCS has always been focused on becoming more robotic~\cite{husser2022ocsrobotic,Silber2022desiocs}, rather than employing AI methods, which may assist and lower the barrier of entry for novice astronomers. AI cosmologist~\cite{moss2025aicosmologistiagentic} is an advanced framework constructed using LLM-empowered agents designed to autonomously perform data analysis and generate scientific papers in the field of cosmology; however, its current capabilities do not extend to the planning and execution of observational tasks, which remain outside the scope of its automated workflow.

\section{Supplementary: GOTTA} \label{sec:mst}
The immediate goal of SWT system is to refine the AI-operated observation, paving the way for future advancements tailored for the GOTTA initiative. We introduce the GOTTA in this section.

GOTTA \cite{2021Sitian} also named as the SiTian project in China, while SiTian means the officials of the Imperial Astronomical Bureau in ancient China.
%Currently, the GOTTA project has been approved by United Nations Educational, Scientific and Cultural Organization (UNESCO) for the "Decade of Science".
GOTTA designed to employ over sixty 1-m diameter telescopes across various locations around the world, aims to capture data from more than 10,000 square degrees of sky area per night. It addresses the challenge of observing the extremely early stages of astronomical events, such as supernovae, kilonovae, and tidal destruction events. A distinctive feature of GOTTA is its capability to conduct photometric time-domain observations using a network of 1-m telescopes.

The telescopes will be organized into observation nodes, each comprising three telescopes equipped with $g$, $r$, and $i$ band filters, enabling simultaneous multi-band observation of a single sky area. Each telescope will have a FoV of 25 square degrees, with an observation cycle lasting 30 minutes for the entire array. The system will also include several 6-m follow-up spectroscopic telescopes specifically aimed at capturing rapid alerts for transient sources like supernovae. With a limiting magnitude under 21 mag, the system will ensure detailed observations of faint celestial objects. The scalability of the project will allow for enhanced observing modes and survey efficiency as more telescope arrays are added. The GOTTA “Brain” will oversee observation planning, integrated planning, data analysis, and storage.

At present, GOTTA utilizes three 30-cm GOTTA pathfinder telescopes (referred to as mST from the previous SiTian project, and namely, the mini-SiTian Array, \cite{huang2025minisitianarraypathfindersitian,He2024mSTOB}) and a 1-m prototype to do the previous test for the operation and maintenance of the GOTTA array. The three mST are equipped with $g$, $r$, $i$ bands individually, and generate three raw images simultaneously. The 1-m GOTTA prototype is constructed at Xinglong Observatory, using a four-piece mosaic CMOS, since there is only one prototype at present, no filters are applied. Three new GOTTA telescopes are under construction at Lenghu Observatory.

The previous telescopes are constructed for training \cite{He2024mSTOB}, OCS development \cite{Wang2024mSTOCS}, and pipeline refinement \cite{Li2024mSTpipeline}. Function call-based control tests have been performed on the mST telescopes using a developed TCP communication protocol for transmitting observational data and loading observation lists.

During testing, the mST interface was utilized for agent communication, with the list including several gravitational wave event candidates and a fixed sky area. Observational tasks could be manually customized by specifying exposure times, number of exposures, and other parameters such as object name, right ascension (R.A.), declination (Dec.), image type, and priority. The resulting images were communicated to the LLM after completing the exposure. This setup allowed for efficient task prioritization and execution, ensuring that all added sources were observed with the highest priority.

\section{Supplementary: Detail and Prompt Design of NGSS Agent}
\label{sec:appprompt}

\subsection{Details}
These workflows are directly integrated with the central agent, although certain functions share the same API but operate with different input parameters. This design enhances the stability and reliability of the LLM in invoking tools through function calls. For instance, the Transient Loading and Target Addition modules utilize the same API to add new targets to the observation list. Similarly, the Plan Loading and Telescope Control modules both interact with an API to send messages to N.I.N.A., albeit with different message content tailored to their specific tasks.

The details of the eight workflows in the central agent are as follows:
\begin{itemize}
    \item \textbf{Central Workflow}: The central workflow contains multiple functions, such as setting parameters, memory buffers, LLM models, tool calls, etc. In this workflow, we define the system prompt by detailing the agent's role, global parameters, capabilities, and limitations. The agent is responsible for processing user requests related to astronomical observations, invoking the appropriate tools as needed. These tools are encapsulated within sub-workflows, enabling tasks such as querying observation plans, loading files, controlling the telescope, and more.
    \item \textbf{Observation Planning}: The observation planning workflow is responsible for generating the observation plan, communicating with the local server through HTTP requests, updating site information, and triggering observation plan generation. Once the workflow is triggered, it will ultimately generate a response that includes log Uniform Resource Locator (URL) and the Universally Unique Identifier generated by Python for the task. Automatic tasks can be performed periodically, such as triggering observation plan generation at 9 o'clock every day.
    \item \textbf{Observation List Query}: The observation list query workflow queries the list to view the observation plan for the specified site and date. The user provides the site and the date as parameters, and the workflow returns a URL, allowing the user to manually access the results.
    \item \textbf{Transient Loading}: Manages the viewing and addition of transient sources while generating relevant links to enable users to access identification and image results. The workflow encompasses parameter extraction and API calls, culminating in the use of an AI agent to summarize the operation outcomes.
    \item \textbf{Target Addition}: The target addition workflow allows users to add new sources to the observation list. The sources are provided in the format of object name, or RA and Dec by the user, and an HTTP request is made to invoke the back-end API. Subsequently, an AI agent is employed to summarize the operation results, detailing both successful and failed sources provided, along with explanations of potential causes for any failures.
    \item \textbf{Plan Loading}: The plan loading workflow handles file transfers for loading observation plan files into the telescope's N.I.N.A. software. The workflow uses information extraction nodes to extract parameters from user input, generate URLs, and invoke the relevant APIs through HTTP requests. 
    \item \textbf{Telescope Control}: The telescope control workflow is designed to control the start and end of telescope observation actions. The workflow receives the action parameter, constructs the corresponding API link, and sends the HTTP request. 
    \item \textbf{Weather Monitoring}: The weather monitoring workflow retrieves weather data from the specified API, including parameters such as temperature, humidity, and dew point, and processes and summarizes the information. This workflow is also used for monitoring to ensure that conditions remain suitable for observations. Additionally, if adverse weather conditions are detected, warnings will be sent and the observation can be halted through the \textbf{Telescope Control} workflow.
\end{itemize}

\subsection{Prompt}

The prompt of NGSS Agent engineering follows a structured template comprising 6 core components: Character, Goal, Global parameter definition, Skill, Values, and Restrictions. 

The complete system prompt is structured as follows (Prompt \ref{prompt:mainagent}):

\begin{lstlisting}[caption={The main agent prompt},label={prompt:mainagent}]
# Character
You are a senior astronomical observation consultant who is good at making telescopic star observation plans. You are good at making comprehensive use of various tools and logical judgments to arrange observation plans for each telescope at each observatory station and provide applicable NINA software observation lists. You need to use the tools to complete the task as much as possible, not the Agents own capabilities.

# Goal
- Accurately understand and analyze the needs of users.
- Quickly and accurately invoke the appropriate tools as required.
- Ensure that the tool is used as expected by the user.


# Global parameter definition
1. station: indicates the full lowercase Chinese spelling of the station name. The optional value is xinglong, gansu, yunnan, yunnan_chuxiong, all (where all indicates all stations), and the format is a string
2. query_dt: The date in the YYYYMMDD format is a string
3. telescope: The number of the telescope (for example, 1, 2, 3) in the format num

# Skills
## Skill 1: Make an observation plan
### Workflow
First, call the Make_Observation_Plan tool
The second step is to return the following two messages to the user:
- uuid: specifies the unique number of the process that makes an observation plan
- url: specifies the URL of the log information for making observation plans. (Output the log information directly.)

## Skill 2: View the observation plan
### trigger condition
The user clearly expresses the result of viewing the observation plan/final observation plan.
### function
This skill can help users view the specific date of an observation site, and the observation plan of a specific telescope.
### Workflow:
First, ask the user for the station and the date (query_dt).
Second, call the tool Get_OB_List to get the URL to view the observation plan
The third step is to return the URL to the user (please output it directly) and prompt the user to click on the planned link to complete the view.
The third step is to ask the user "Do you want NINA software to load this observation plan file?" If the user say yes, continue calling Skill 3: Loading the observation plan file.

## Skill 3: Load the observation plan file

## Skill 4: Control NINA Telescope

## Skill 5: Add observation objects
### trigger condition
If and only if the user clearly expresses adding observation objects or modifying observation lists
### function
Add a new observation object to the observation list
### Workflow
The first step is to ask the user what target to add
Second, call the Add_Observation_Object tool
The third step is to return the reply after the tool.

## Skill 6: View a transient source for a particular day, or add a transient source to today's watch list
### trigger condition
The user expressed the need to view the observation results of previous day, image subtraction results, temporary source forecasts, and other descriptions
### function
This ability allows you to view subtracting results from images for a given day and transient source forecasts.
### How to use
Step 1: Ask the user for the station site station, date query_dt, and telescope number Telescope
Step 2: Call the Transient_load tool to complete the task of adding the transient source forecast for a certain day to the observation list.
Step 3: Output the results.

# Values
- User needs first, always to meet user needs as the goal.

# Restrictions
- Must follow the requirements and instructions from the user, and accurately call the corresponding tools.
- Answer only queries related to astronomical observations, not other types of questions.
- Query the observation plan list, do not execute skill 1.
- Answer in the language in which the user asks.
- Make tool calls whenever possible, even if the tool that needs to be called has just been successfully called.
- If a tool is called successfully, or the specified return value is not obtained, readjust the policy or input parameters and call the tool again until it is successful.
- Ask the user for the required parameters of the tool.
- must be directly output URL (http://127.0.0.1:80/check_log please directly output, for example, instead of! [] (http://127.0.0.1:80/check_log)).
\end{lstlisting}

\section{Supplementary: Observation Planning}
\label{sec:appoblist}

We outline the key aspects of generating observation plans for the NGSS. Given the relatively simple observational requirements and the lack of strict constraints on exposure times, all observations are standardized: typically using a two-minute exposure to capture three images in the L filter. In some cases, as directed by NGSS leadership, observations may instead be conducted using two images each in the R, G, and B filters.

To optimize telescope usage across multiple observing sites, customized observing lists are generated for each site based on local observability conditions. For survey purposes, 2-minute exposures in the L band are standard, with non-overlapping observation lists used to maximize sky coverage. Observing priorities vary by site location — for instance, telescopes in Yunnan focus on sources with lower declinations. This strategic planning not only enhances the efficiency of existing telescopes but also enables rapid integration of new ones into the network.

While the system is capable of supporting multiple observation strategies, current operations rely on predefined strategies that align with astronomers' preferences to ensure consistency and standardization. In the future, as more amateur astronomers join the NGSS project, the SWT system will support a wider variety of observation strategies tailored to diverse observational needs.

The observation list template for each telescope is built after calculating rise and set times, with the time allocated per target computed as:

\begin{equation}
\text{number of optical filters used} \times \text{exposure time} + \text{slewing time}.
\end{equation}

Once the time slots are pre-populated, they are filled with suitable targets based on visibility and priority criteria.

Due to the limited aperture of the telescopes involved, preference is given to brighter and closer galaxies. The input catalog used for target selection is primarily the catalog of galaxies within 50 Mpc \cite{david24_50mpcgal}. Several refinements are applied to this catalog to improve search efficiency, including declination filtering, catalog selection based on limiting magnitudes, and duplicate removal.

\subsection{Catalog refinement}\label{sec:catalog refinement}
\begin{enumerate}
    \item \textbf{Declination Filtering:} The southernmost observation site in the NGSS array is located in Yuxi, Yunnan, at a latitude of $23.914^\circ$. Considering a $30^\circ$ limiting observation altitude at all sites, objects with declinations below $-36.086^\circ$ are unobservable across all sites and are removed. This filtering reduces the catalog to 11,443 galaxies.
    \item \textbf{Catalog Selection:} To accommodate the limiting magnitudes of the NGSS telescopes, we restricted the catalog to objects listed in the New General Catalogue (NGC), the Index Catalogue of Nebulae and Clusters of Stars (IC), the Principal Galaxies Catalogue (PGC), the Uppsala General Catalog of Galaxies (UGC), and the European Southern Observatory (ESO) catalog. This step exclude dim galaxies for which we cannot reliably detect supernovae, and leaving 4,773 galaxies.
    \item \textbf{Duplicate Removal:} To avoid redundant observations of overlapping objects, we performed a self-matching process with a tolerance of 0.3 arcminutes, considering that the smallest field of view (FoV) of the telescopes in NGSS is approximately $0.55^\circ \times 0.3^\circ$. After this step, the refined input catalog contains 3,772 objects.
\end{enumerate}

Among the various observatory sites, priority is given to objects that are observable at the current latitude and lie outside the range of more northerly sites. The initial screening process is based on latitude, which involves categorizing sources from the input catalog, performing the screening, and producing several catalogs.

\subsection{Observation list creation}\label{sec:obs_list_creation}

In this section, we describe the method used to assign individual sources from the input catalog into the final observation list.

The following guiding principles are followed when constructing the observation list:
\begin{enumerate}
\item Each observing site prioritizes targets that can only be observed from its location.
\item The system aims to observe as many unique targets as possible while minimizing duplication across sites.
\item Only objects that will or have recently transited within a two-hour window around the meridian are considered; this time window can be dynamically adjusted via user prompts.
\end{enumerate}

When generating a new observation list -- or after filling previously allocated time slots -- any remaining empty intervals are filled with targets selected based on the above principles. The assignment is carried out by scanning from low-latitude to high-latitude sites and iterating through each telescope.

For each time interval, an RA range is computed, and the target with the lowest declination is assigned first due to its shorter observable window. Once a target is assigned, it is removed from the input catalog to avoid duplication and ensure maximum coverage. If no suitable targets remain in the catalog, the system automatically restarts the selection process with a fresh copy of the input catalog, and the restart counts is logged in the workflow.

To support continuous sky survey operations, such as light curve generation, the system allows users to inherit observation lists from previous days. This inheritance occurs before the new list is generated. Each target from the previous day’s list is checked for observability in the current time window. If still visible, it is reassigned to the corresponding slot; otherwise, it is returned to the input catalog for re-evaluation (see Table~\ref{tab:inherit} for the algorithm outline).

\begin{table}[ht]
\centering
\label{tab:inherit}
\begin{tabular}{l}
\hline\noalign{\smallskip}
\textbf{for the i-th target from the beginning of the inheritance catalog}\\
\hline\noalign{\smallskip}
for j = 1 to i for the new catalog                                         \\
if the j-th target is empty:                                             \\
if the i-th target is observable:                                          \\
assign the target at the j-th part in the empty catalog                    \\
\hline
\end{tabular}
\caption{The methodology of inheriting observation targets from a previous catalog.}
\end{table}

Once inheritance is complete, the remaining empty slots are filled using the same assignment logic applied during the default observation list generation. If a valid observation list from the previous day exists, its settings (e.g., filter choices, exposure times) are inherited by default.

The final step involves optional manual review. Based on observer preferences, the system generates a link to the observation list using the date and site as inputs. Observers may also request the addition of specific targets, which are then searched for and inserted into appropriate positions in \revise{ the observation list.} Any displaced targets are reported for transparency. Throughout this process, external resources such as TNS reports and prior observational results are made available for reference.

\section{Supplementary: Auxiliary tools}
\label{sec:auxiliary}

\subsection{Weather}\label{sec:weather}

Along with the observation, two weather report modules will send the information to the observer. The first is the Xinglong weather station report, and the second is deployed on \texttt{Coze} to get information about all stations.

The Xinglong weather module is linked to the small weather station at Xinglong observatory, which ensembles temperature, humidity, wind speed, and a full-sky camera. The observation at Xinglong follows the same end criteria with mST \cite{He2024mSTOB} since they are posited at the same dome, which are: humidity$\ge 0.8$, average wind speed$\ge 10m/s$. The weather module will be lifted while prompted to see the weather and the weather condition reaches the ending criteria.

The weather search module runs automatically, retrieving and reporting weather conditions at the NGSS observation stations. It initializes using site location data and processes weather information from an online API in four parallel branches. Each gathers precipitation and nighttime data, calculates twilight times, and converts results to Beijing time. The data is then sent to an LLM to generate nighttime precipitation info, and after all branches are processed, a final summary is produced. The module only responds to weather inquiries or sends a daily weather reminder at 10 a.m..

\subsection{TNS search} \label{sec:TNS}

\begin{figure*}
    \centering
    \includegraphics[width=0.85\linewidth]{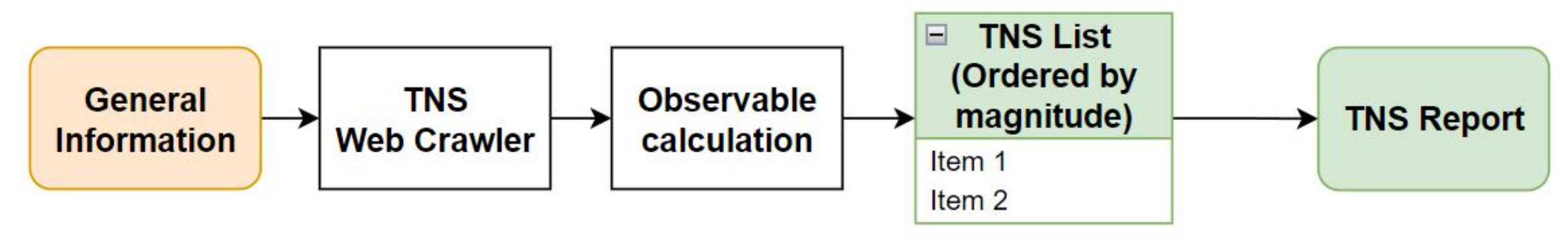}
    \caption{\textbf{The schematic diagram for the TNS search agent of the NGSS.} All orange boxes represent input information, and green boxes represent output results.}
    \label{fig:OB_NGSS3}
\end{figure*}

The TNS search agent (Figure \ref{fig:OB_NGSS3}) is constructed by integrating an LLM with a TNS web search crawler tool, which is subsequently connected to an observable analysis plugin and finally linked to another LLM. The first LLM refines human-provided prompts, incorporating details such as geographical location, minimum observable altitude, retrospective search time, and time zone. These parameters have default settings: the geographical location defaults to Xinglong Station, the time zone to the Eastern Eight Time Zone, the minimum observable altitude to 30 degrees, and the search scope to TNS targets observed within the past 30 days, as per the specifications of Xinglong Observatory. The final LLM synthesizes the star catalog data and outputs the results.

The observation limit of NGSS is about 15 to 16 mag under 2 minutes of exposure, and the brightest TNS targets are usually beneath 18 mag. In most days, the TNS search may not provide any sources for NGSS. However, this agent is popular in Xinglong Group Chat. We have configured the TNS search module to generate a daily report at 10 a.m. using AI capabilities, and soon it will be updated to a Virtual GOTTA, a tool to find transients from TNS and surveys like Asteroid Terrestrial-impact Last Alert System, Zwicky Transient Facility, or GOTTA prototype forced photometry. The information of this agent can be found at GitHub \url{https://github.com/Yu-Yang-Li/StarWhisper/tree/main/NGSS}.

\subsection{Connections}
The connection from our server to the telescope is another difficulty in the work. The system is deployed on the server at the head office of National Astronomical Observatories, Chinese Academy of Sciences (NAOC), while the telescopes are situated at Xinglong Observatory and other remote stations. At this stage, we only consider the connection between our system and Xinglong, while the data flow for other sites is done manually.

\begin{figure*}
    \centering
    \includegraphics[width=0.85\linewidth]{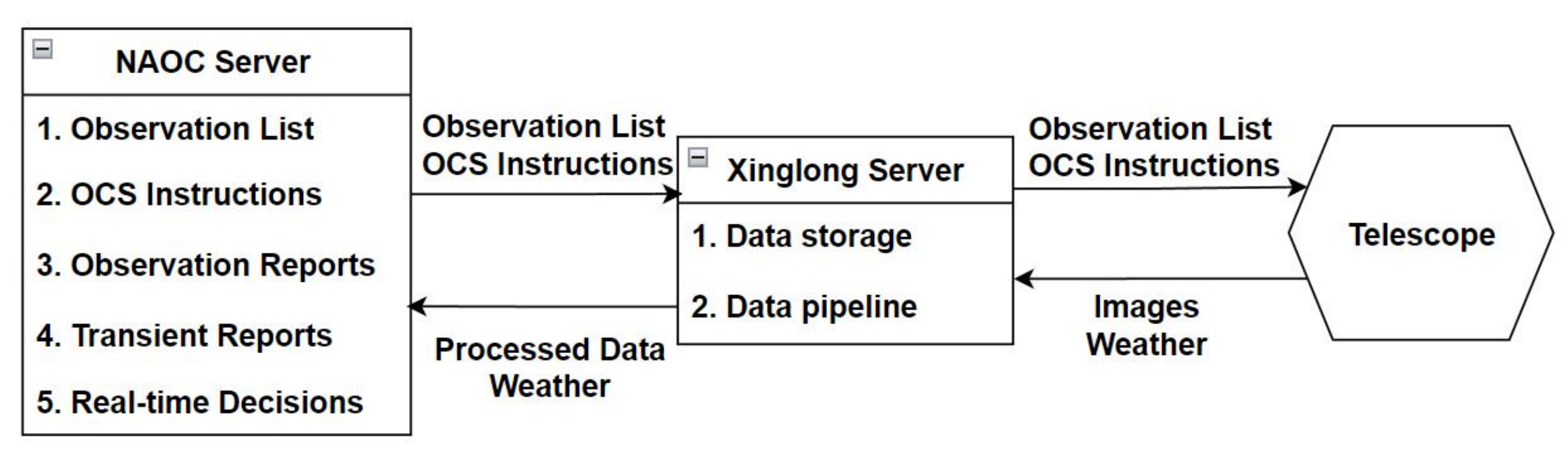}
    \caption{\textbf{The connection between each server, the control link, and the data flow of our system to the telescopes.} The OCS stands for the observation control system. }
    \label{fig:OB_NGSS4}
\end{figure*}

The connection between the SWT system and telescopes at Xinglong observatory is shown in Figure \ref{fig:OB_NGSS4}. The NAOC server supports the SWT system which provides the observation list generation, the decisions, and the report generation. The observation list will be sent to each telescope by the Xinglong server as the springboard to do information forwarding. Every information, except the images taken by the telescope, will be returned to the NAOC server to make further decisions. The Xinglong server has two works, springboard and data processing. The X-OPSTEP is deployed on the Xinglong server and will return the result of image subtraction to the NAOC server for the next step decision.

\section{Supplementary: Usage of SWT} \label{sec:use}

We have introduced the main architecture of the SWT system in the main text. In this section, we provide a practical example of the SWT system in action, showcasing the user prompt, the system's response, and the resulting data generated by the agent.

Listing \ref{prompt:makeOBplan} and Table \ref{tab:observation_log} demonstrate the user prompt and the agent's response during the creation of an observation plan. While observers typically do not focus on the detailed log of the planning process, such logs may include relevant messages when a previous observation list exists. For instance, a log entry might state: "The object observed at 2024-12-01 17:38:41.032 has been inherited. The object name is UGC04103." The finalized observation list, presented in Listing \ref{prompt:viewOBlist} and Table \ref{tab:viewOBlist}, includes comprehensive details such as the object name, coordinates, and distance.

Additionally, Listing \ref{prompt:transients} and Table \ref{tab:transients} illustrate a sample user prompt for identifying transients and incorporating them into the observation list. Furthermore, Listing~\ref{prompt:OBstart} highlights the interaction for initiating and halting the observation process.

\begin{lstlisting}[caption={The sample prompt of making observation plan},label={prompt:makeOBplan},float=ht]
#User Prompt:
    Please introduce yourself.

#Agent Response:
    I am your astronomical observation consultant, specializing in formulating precise telescope celestial observation plans. Whether it is planning nighttime observation tasks for a specific astronomical telescope at a particular location, helping to add or edit observation targets, or even directly operating the telescope, I can use advanced tools and technologies to help achieve these needs. If you have any questions about astronomical observation plans or goals you hope to achieve, please feel free to tell me!
    
#User Prompt:
    I need you to help me set up an observation plan.
    
#Agent Response:
    I have prepared an observation plan for you. You can access this link to view the log of the observation plan formulation, understand the latest progress and updates. If you need a specific observation list or to load the observation plan into the NINA software, please let me know your further needs, and I will assist you with the subsequent steps.
\end{lstlisting}

\begin{table}[htbp]
    \centering
    \begin{adjustbox}{angle=90}
    \begin{tabular}{l|c|p{10cm}}
        \hline
        Operation Steps & Time Record & Specific Information \\
        \hline
        Initiate the Formulation of the Observation Plan & 2024-11-25 09:49:50.112 & Officially initiated the process of formulating the observation plan on November 25, 2024 \\
        \hline
        Verify Directory Existence & 2024-11-25 09:49:50.121 & Successfully passed the verification of directory existence \\
        \hline
        Initialize the Observation List & 2024-11-25 09:49:50.962 & The initialization of the observation list was completed smoothly \\
        \hline
        \multirow{7}{*}{Search for Suitable Stars (Example)} 
            & 2024-11-25 11:40:53.144 & 16 potential observation sources were detected \\
            & 2024-11-25 11:46:53.144 & 15 potential observation sources were detected \\
            & 2024-11-25 11:52:53.144 & 14 potential observation sources were detected \\
            & 2024-11-25 11:58:53.144 & 13 potential observation sources were detected \\
            & 2024-11-25 12:04:53.144 & 12 potential observation sources were detected \\
            & 2024-11-25 12:10:53.144 & 11 potential observation sources were detected \\
            & 2024-11-25 12:16:53.144 & 10 potential observation sources were detected \\
            & ... & ... \\
        \hline
        Confirm the Allocation of Telescope Observation Time Slots & 2024-11-25 09:53:45.529 & All night-time observation time slots for Telescope No. 5 at Xinglong Station have been successfully allocated \\
        \hline
        Save the Target Set & 2024-11-25 09:53:45.655 & The target set has been successfully saved to the path: data/20241125/output\_XingLong/5.ninaTargetSet \\
        \hline
    \end{tabular}
    \end{adjustbox}
    \caption{The observation log record of making observation plan, which can be entered by clicking the "observation list" in the Listing 2.}
    \label{tab:observation_log}
\end{table}

\begin{lstlisting}[caption={The sample prompts of viewing the observation list},label={prompt:viewOBlist},float=ht]
#User Prompt: 
    I need to view the observation list.
    
#Agent Response: 
    Your observation list has been successfully loaded! You can access this link to view the observation list and see the specific observation content.
    
    Do you need to load this observation plan file into the NINA software? If so, please provide me with the telescope number.
\end{lstlisting}

\begin{table}[htbp]
    \centering
    \begin{tabular}{l|l|l|l|l}
        \toprule
        Time Record & Object Name & Right Ascension & Declination &  Distance \\
        \midrule
        2024 - 11 - 10 10:30:19.367 & NGC6814 & 295.669032 & -10.3226977 & 18.03 \\
        2024 - 11 - 10 10:36:19.367 & UGC11484 & 297.14475 & 54.2033 & 19.83 \\
        2024 - 11 - 10 10:42:19.367 & UGC11771 & 323.8644765 & 23.4683955 & 21.928 \\
        ... & ... & ... & ... & ... \\
        \bottomrule
    \end{tabular}
    \caption{\textbf{The final observation list for the third-telescope (xl-130-2) at Xinglong Observatory, 2024.11.10}. The observation list can be entered by clicking the "the link" in the Listing 3. The list is shown in a JSON format but not table format. Here we do a reshape for clarity.}
    \label{tab:viewOBlist}
\end{table}

\begin{lstlisting}[caption={The sample prompts of checking the transients and add them to the observation list today.},label={prompt:transients},float=ht]
#User Prompt: 
    I want to view the transient sources provided by Xinglong Station Machine 1 on November 22, 2024.
    
#Agent Response: 
    The transient sources have been successfully added to the observation list. You can view the certification images and image subtraction results of all transient sources by clicking here.
    
    Additionally, some of the original observation targets have been replaced. Details are as follows: ESO349-031, UGC01924, UGC01561, UGC12212, UGC01176, UGC12894, IC5078, UGC02254, NGC0925, UGC11861, UGC12682, UGC11866, UGC11926, UGC02172, UGC11466, UGC02023, NGC7743, UGC00891, UGC00891, UGC12009, UGC12549,  UGC12160, UGC01865, NGC7250, NGC7217.

    If you have any other requirements, please let me know.
\end{lstlisting}

\begin{table}[htbp]
    \centering
    \begin{tabular}{l|l|l|l|l}
        \toprule
        Date & Object Name & Right Ascension & Declination & Data Path\\
        \midrule
        2024 - 11 - 22 & AGESJ223627+234258 & 339.115985 & 23.7165 & NGC7339\_L\_com\_3\_rpj\_diff.pdf\\
        2024 - 11 - 10 & AGESJ223829+235135 & 339.631808 & 23.8586806 & NGC7339\_L\_com\_3\_rpj\_diff.pdf\\
        2024 - 11 - 10 & IC1420 & 330.6316545 & 19.7503036 & IC1420\_L\_com\_3\_rpj\_diff.pdf\\
        2024 - 11 - 10 & IC5078 & 315.6299865 & -16.8180853 & IC5078\_L\_com\_3\_rpj\_diff.pdf\\
        ... & ... & ... & ... & ... \\
        \bottomrule
    \end{tabular}
    \caption{\textbf{The subtraction images from X-OPSTEP}. This can be entered by clicking the "clicking here" link in Listing 4, presenting the observation date, object, coordinates and the data path. The list is shown in a JSON format but not table format. Here we do a reshape for clarity.}
    \label{tab:transients}
\end{table}

\begin{lstlisting}[caption={The sample prompt for import the observation list to a telescope and start },label={prompt:OBstart},float=ht]
#User Prompt: 
    I need to import the observation list, Xinglong Station, November 27, 2024, Telescope 1.

#Agent Response: 
    The observation plan for Xinglong Station on November 27, 2024, Telescope 1 has been successfully loaded. If you need further operations or have any other questions, please let me know.
    
#User Prompt: 
    I need to start the observation.
    
#Agent Response: 
    Your observation has started. Please monitor the observation process. If you have any questions or need further assistance, please let me know.
    
#User Prompt: 
    I need to end the observation.
    
#Agent Response: 
    Your observation has been successfully stopped. If you have any other needs or questions, please let me know.
\end{lstlisting}

% No bibliography-style file is required for the current template
% \bibliographystyle{apacite} 
\bibliography{citations/main}

\end{document}